# Assessment of waterfront office redevelopment plan on optimal building energy demand and rooftop photovoltaics for urban decarbonization

Younghun Choi[1], Takuro Kobashi[1,*], Yoshiki Yamagata[1], and Akito Murayama[2]

[1]Center for Global Environment Research, National Institute for Environment Studies, 16-2 Onogawa, Tsukuba, Ibaraki 305- 8056, Japan

[2]Department of Urban Engineering, School of Engineering, The University of Tokyo; 7-3-1 Hongo, Bunkyo-ku, Tokyo 113-8656 Japan

*Corresponding author: kobashi.takuro@nies.go.jp;

**Abstract**

Designing waterfront redevelopment generally focuses on attractiveness, leisure, and beauty, resulting in various types of building and block shapes with limited considerations on environmental aspects. However, increasing climate change impacts necessitate these buildings to be sustainable, resilient, and zero $CO_2$ emissions. By producing five scenarios (plus existing buildings) with constant floor areas, we investigated how building and district form with building integrated photovoltaics (BIPV) affect energy consumption and production, self-sufficiency, $CO_2$ emission, and energy costs in the context of waterfront redevelopment in Tokyo. From estimated hourly electricity demands of the buildings, techno-economic analyses are conducted for rooftop PV systems for 2018 and 2030 with declining costs of rooftop PV systems. We found that environmental building designs with rooftop PV system are increasingly economical in Tokyo with $CO_2$ emission reduction of 2-9% that depends on rooftop sizes. Payback periods drop from 14 years in 2018 to 6 years in 2030. Toward net-zero $CO_2$ emissions by 2050, immediate actions are necessary to install rooftop PVs on existing and new buildings with energy efficiency improvements by construction industry and building owners. To facilitate such actions, national and local governments need to adopt appropriate policies.

1. Introduction

1.1. Urban decarbonization

About 75 % of global power consumption and 60-70% of greenhouse gas emissions originate from cities [1,2]. However, as the center of economic competitiveness and innovation, cities are also the sources of solutions [2,3]. Smart city is one of the necessary ingredients to urban sustainability contributing on recent urban challenges such as rapid expansion of urban population and decarbonization. Increasing digitization, development of Information and Communication Technology (ICT) and artificial intelligence (AI) is expected to play substantial roles on the development of decentralized urban power systems. In addition, declining costs of PV systems and EVs with increasingly tighter regulations are rapidly introducing these technologies into urban energy systems, which are integrated by the smart city technologies as distributed energy resources (DER) [4]. Studies indicated that rooftop PVs plus EVs as batteries can play substantial roles on urban decarbonization supplying up to 95% of affordable $CO_2$-free electricity to urban dwellers in nine Japanese cities known as the SolarEV City concept [5,6].

The Government of Japan announced that Japan aims to reach net-zero emission by 2050. Therefore, it is critical that all urban planning processes are to be assessed for future zero-emission. As Japan is constituted in four main islands with long coastlines, waterfront redevelopments are one of the higher priority policy options for many local governments to increase life quality of citizens and to attract tourists. As a consequence, many redevelopments of river or coastal sides are taking place [7–9] with benefits to improve economic values, environmental conditions, transport and social services, economic investment opportunities on currently degraded areas. At the time of rapid energy transition toward net-zero $CO_2$ emission, this waterfront redevelopment planning must also integrate energy efficient building, block design, renewable energies such as tidal power, hydroelectric power, and solar power for their energy demands.

Energy demands for office buildings are created for various services such as lighting, space cooling and heating, office appliances, elevator, etc. Space heating and cooling demands (e.g., about 28% of the total office building energy demand in Japan [10]) are controlled by various factors such as building wall materials, efficiency of heating, ventilation, and air conditioning (HVAC) system, building shapes, and influence of shades by neighboring buildings. Therefore, to achieve efficient building energy systems, it is necessary to conduct energy assessments in the early planning phase of redevelopment with proper tools and methods [11–14]. Also, retrofitting existing buildings needs to be considered for a rapid reduction of $CO_2$ emission to reach net-zero emission by 2050 [15,16]. Expected rapid developments of PV technologies for the coming decades in terms of costs, efficiency, weights, and design, will provide unprecedented opportunities for these measures to be effective and beneficial to building owners.

1.2. Urban building energy modeling and techno-economic analysis

Urban building energy modeling (UBEM) with three-dimensional (3D) representation are rapidly developing, and more and more applied to assess sustainable urban building forms [13,17–19]. For example, "Rhinoceros 3D" is a computer 3D graphics for computer-aided design (CAD), and its plug-in Grasshopper is a visual programming environment. "Grasshopper" hosts various energy modeling tools such as Ladybugs and Honeybee, which further connect with a well-known building energy modeling tool such as "EnergyPlus". The analysis can be made for a building or building blocks in various resolution in time and space. As often hourly building energy demand is not publicly available owing to privacy, etc., these models are important for the assessments on the viability of variable renewable energies (VREs) such as BIPV in urban environments.

The tools have been utilized for various applications. Natanian et al. [12] analyzed various nearly zero energy building and district types between courtyard, scatter, slab, high-rise, and courtyard in the hot/dry climate of Mediterranean. They found the courtyard typology performs to be the best option in terms of energy balance, but with less optimal performance in daylight utilization. Then, also Natanian et al. [20] introduced an energy and environmental quality evaluation workflow. Zhang et al. [21] compared energy demand and solar potentials of different block types in the hot and humid climate of Singapore. They found solar energy harvesting amount can increase up to 200% depending on block types with other variables constant except morphology. Chang et al. [14] investigated relationship between design parameters and urban performance parameters such as energy demands, solar harvesting potential, and sky view factor for university campus design in Shenzhen, China. They applied statistical approaches and identified optimal building coverage ratio and sky view factor.

Actual implementation of renewable energy projects such as BIPVs depends on financial merits in comparison to existing energy systems such as grid electricity [22]. Techno-economic analyses can assess if a renewable energy project is viable considering the costs of technologies, discount rate, project period, degradation, electricity tariff, insolation changes, etc. [23]. As the cost of PV systems is expected to drop further [24], the viability of PV projects also improves significantly in the coming decades, increasing potentials of rooftops PVs [25]. Many studies have been conducted to test viability of rooftop PV systems coupled with battery and EV as battery for households [26–28]. For example, Lang et al. investigated residential and commercial buildings for viability using techno-economic analyses for Germany, Switzerland, and Austria [29]. They found the rooftop PVs are already attractive to many buildings without subsidies. However, few studies have investigated impacts of building and block design for waterfront office redevelopment on energy demand, rooftop PV generation, $CO_2$ emissions, considering declining costs of PV systems from 2018 to 2030.

In this study, we conducted environmental and energy analyses for waterfront office building redevelopment in Shinagawa, Tokyo, Japan as a test site (Fig. 1). We produced five scenarios with different building and district forms (scenario 1-5) in comparison to existing buildings (scenario 0), which include nine buildings on average. Energy demands of buildings and rooftop PV generation in an hourly resolution are estimated for all the scenarios using "Rhinoceros 3D" and its plug-ins Grasshopper, etc., considering energy balance between in- and out-side of buildings with a weather file for 2018 as an input. Total floor areas of buildings and site area in all the scenarios are set constant for comparison purposes. Then, techno-economic analyses were conducted using System Advisor Model (SAM) [30] to assess the viability of rooftop PV systems for 2018 and 2030 to evaluate impacts of increasingly cheaper rooftop PV systems. Finally, environmental and energy indicators such as $CO_2$ emission, self-sufficiency, self-consumption, and energy sufficiency were evaluated for each scenario as well as financial indicators (net present values (NPV), payback periods, and levelized costs of electricity (LCOE)).

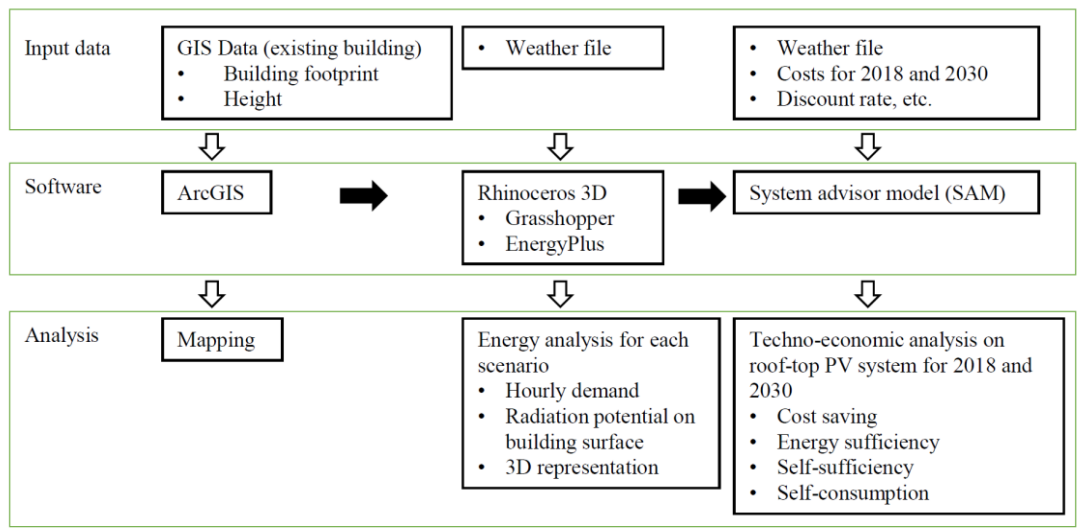

Figure 1. Evaluation workflow for waterfront building environmental energy analyses.

In the following section 2, methodologies of the analyses were presented using Rhinoceros 3D, Grasshopper, and SAM. In the section 3, estimated hourly energy demands for scenarios were presented, and various indicators were calculated and compared between scenarios. The implications of the results were discussed in the section 4. Finally, we summarize and conclude our findings in the section 5.

## 2. Materials and Methods

### 2.1. Shinagawa, Tokyo

The test site, Shinagawa area (35.6° N, 139.7° W) is located near Shinagawa Railway Station in Minato Wards, Tokyo, Japan. The Shinagawa Station is one of the busiest railway stations in Japan with annually 380 thousand users. Land use of Shinagawa is divided by the Shinagawa Station. West side of the station is mainly for residential-oriented mixed-use area, and east side is office/industry-oriented mixed-use area where the test site is located. The test site (Fig. 2) is near harbor along Tokyo Bay with canals going through the middle of the district. Currently, this waterfront area is not actively utilized as a recreation area considering their potentials. Shinagawa experiences maximum daily average temperature of 30 °C in summer and minimum temperature of 0 °C (Fig. 3) with snow fall only occurring a few times a year. Coastal regions of Japan along Pacific Ocean including Shinagawa have generally fine weather in winter, reflected in high quality PV generation but with shorter daytime (Fig. 3).

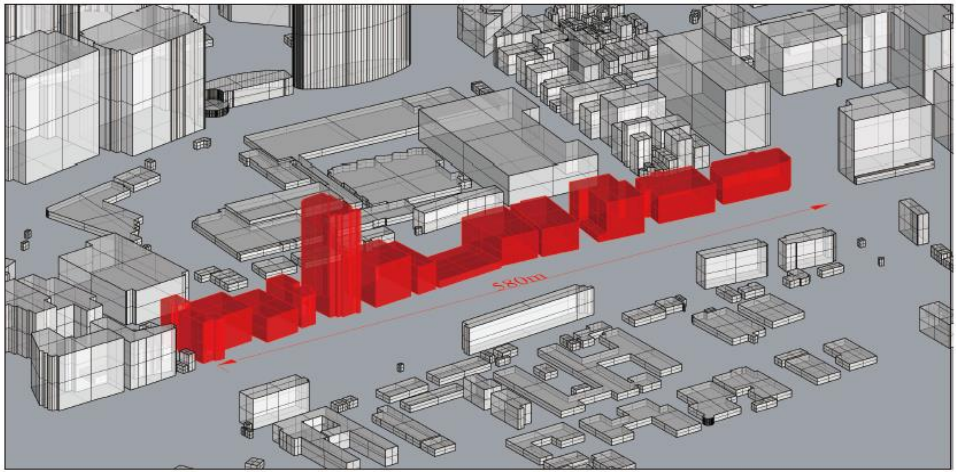

Figure 2. "Existing buildings" with surrounding buildings in Shinagawa area, Tokyo. Red colored buildings were analyzed as existing buildings (scenario zero). It is noted that there is a canal in front of the existing buildings.

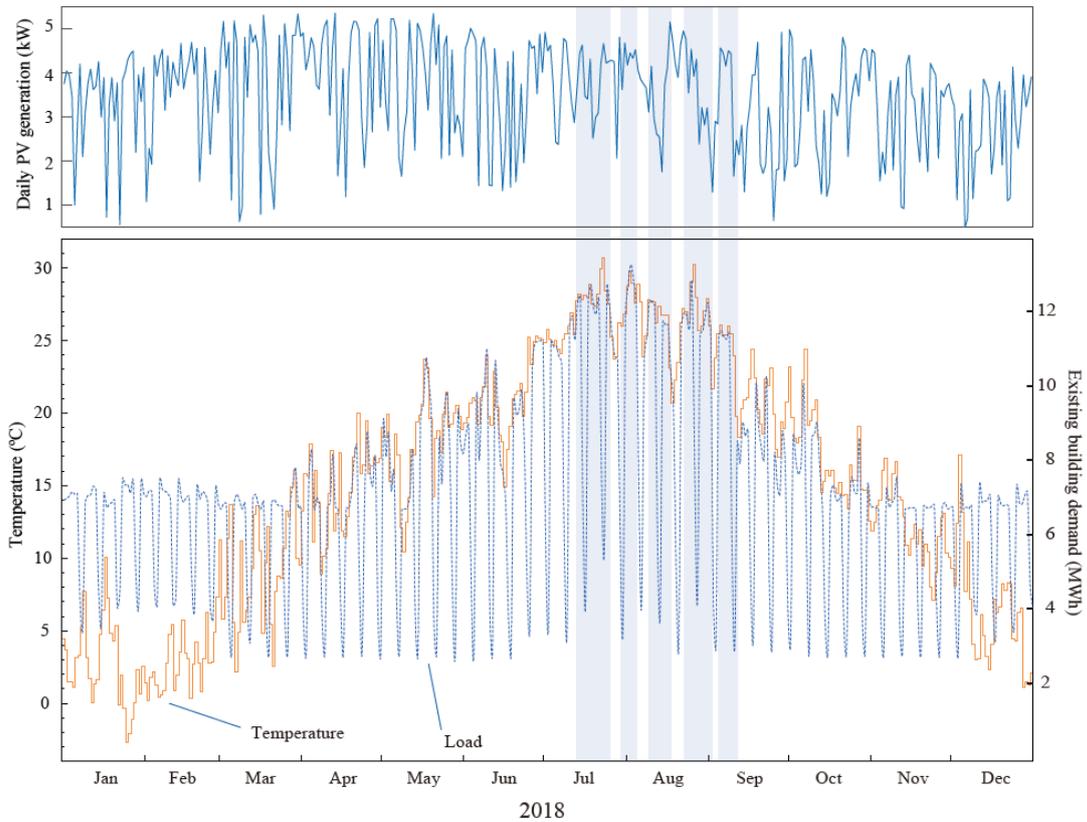

Figure 3. PV generation, daily outside temperature, and estimated demand of existing buildings for the test site in Shinagawa for 2018. Blue bands represent hot days with cooling demand.

## 2.2. Environmental and energy analysis tools

Environmental energy analyses were conducted as in a workflow chart in Figure 1. "ArcGIS", a Geographic Information System (GIS) program, was utilized to create GIS database and mapping [31]. Footprint and height data of buildings were obtained from publicly available dataset [32], and 3D polygons of buildings were produced by extruding foot-print areas with the corresponding heights of buildings (Fig. 2). Then, the data was saved as a shapefile by "ArcGIS". "Rhinoceros 3D" is a 3-dimensional computer aided design (3D CAD) software. "Rhinoceros 3D" and its plugin, "Grasshopper", provide various analysis tools (e.g., Ladybug and Honeybee for energy analyses) for building designers, allowing them to work with independently developed software such as "EnergyPlus", "Radiance", and "Daysim" [33]. "EnergyPlus" is a well-known program for whole building energy analyses developed by US Department of Energy (DOE) [34]. We used these programs to estimate hourly energy demand and PV generation potentials for buildings [14], considering building usage patterns, materials of walls, windows and rooftops, weather, and urban context such as shades of nearby buildings. Accuracy of EnergyPlus has been tested and validated during its on-going development [35].

The shapefile was loaded into "Rhinceros 3D" (Fig. 1). "Grasshopper" provides a platform to build sequences of energy analyses. A weather file (epw) is necessary to estimate hourly load and PV electricity generation. We used "SIREN" to produce a weather file for Shinagawa, Tokyo for 2018 [36]. "EnergyPlus", integrated within Ladybug, analyzes building hourly energy demands (heating, cooling, lighting, and appliances) considering the influences on shading from neighboring building. PV electricity generation was also calculated on the surface of buildings (kWh/m$^2$) in hourly resolution (Fig. 3) also considering shading (Fig. 4). Annual radiation amounts on the surface of the buildings were calculated for each mesh with an average area of 8 m$^2$ (Fig. 4) Maximum annual radiation amount was calculated as 1383 kWh/m$^2$ (Fig. 4). Above-ground-window/wall ratio for north, west, south, and east faced walls were set as 0.4, 0.35, 0.2, 0.15, respectively (Fig. 5). Floor heights were set to 3 m. Space heating and cooling demands were converted to electricity demand by coefficient of performance (COP) with the values of 2.27 and 2.51, respectively [37]. Owing to the rapid development of the building energy analyses tools and increasing users, the analyses between the program are smoothly linked, and results can be readily projected as 3D building representation in Rhinoceros 3D (Fig.5). Grasshopper files for energy and radiation analyses with weather files are available as supplementary files.

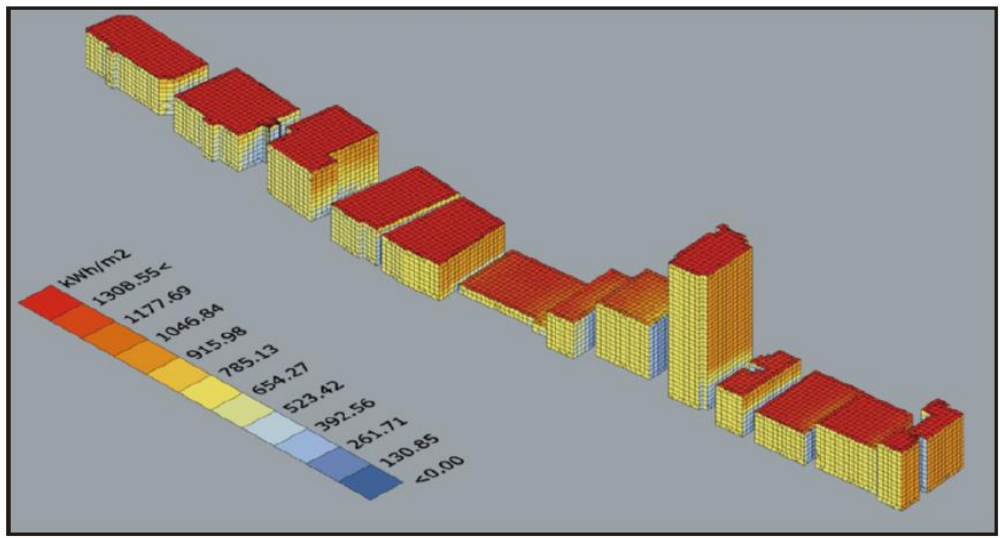

Figure 4. Annual radiation amount (kWh/m$^2$) on the surfaces of the existing buildings calculated by Grasshopper. The direction of increasing radiation in the colorbar represents north.

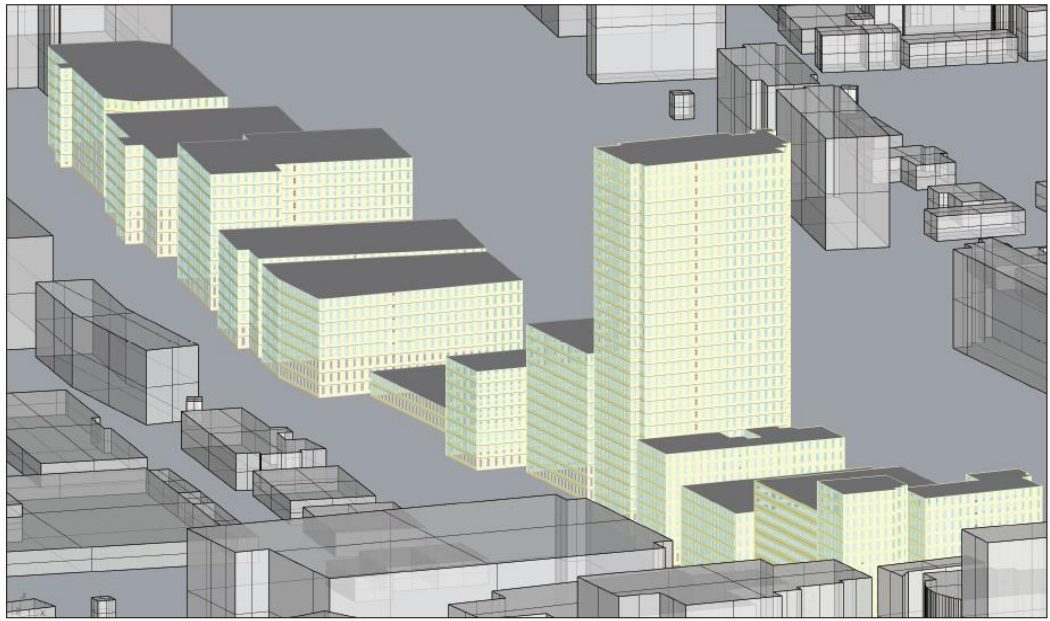

Figure 5. Existing buildings (yellow) with windows for the analyses. Note that ceilings for each floor is not shown.

*2.3. Techno-economic analysis*

Techno-economic analyses evaluate the viability of renewable energy projects such as rooftop PV systems [25,28], comparing with existing energy systems. The analyses consider project periods, discount rates, costs of PV systems, degradation, various energy losses, tariffs, etc. [23]. To investigate the impacts of declining PV system costs, we conducted techno-economic analyses on rooftop PV systems on large office buildings for 2018 and 2030 (Table 1). The methods generally follow those of our earlier studies [5,6]. System Advisor Model (SAM) was used for our analyses. The software is publicly available, and developed by National Renewable Energy Laboratory (NREL) of the U.S. DOE [30]. Hourly energy demands of buildings were estimated using the aforementioned "Rhinoceros 3D". A weather file for 2018 was applied to SAM analyses. We used a project period of 25 years with a discount rate of 3% for the rooftop PV systems. Currency exchange rate of 110 yen/$ was used. Other parameters used for the analyses are listed in Table 1. An electricity tariff price for high-voltage users was utilized for the analysis, which is cheaper than that for low voltage users (households, etc.) in Japan. SAM files with a weather file for the analyses were made available as supplementary files.

Table 1. Parameters used for techno-economic analyses [5]. Small-scale PV system costs are for 2018 (2030), respectively. Maintenance costs for PV system includes inverter replacements.

| Items | 2018 (2030) |
|---|---|
| Small-scale PV system cost ($/kW) | 2.15 (0.88) |
| PV system maintenance cost ($/kW/yr) | 31.4 |
| Electricity to buy ($/kWh) | 0.15 |
| Electricity to sell ($/kWh) | 0.08 |
| PV tilt angle (degree) | 30 |
| Grid emission factor ($kg_{CO2}$/kWh) | 0.455 |

*2.3.1. Net present values*

We used net present values (NPVs) as a primary financial indicator and identified optimal PV capacity for each scenario using a function of SAM, "Parametrics". NPV of a PV project is a sum of discounted annual net saving over the project period including all the costs incurred (e.g., capital, and annual maintenance costs) [30,38].

Therefore, NPV is defined as:

$$NPV(p,t) = \sum_{n=1}^{N} \frac{Cash\ Flow(p,n,t)}{(1+R_d)^n} - System\ Cost(p,t)$$

where

$p$ = PV capacity (kWh)

$t$ = Project first year (year)

$N$ = Project period (year)

$R_d$ = Discount rate

and

*Cash Flow (p, n, t) = Electricity Cost<sub>Base</sub> (n, t) – Electricity Cost<sub>System</sub> (p, n, t)*

*Electricity Cost<sub>base</sub>* and *Electricity Cost<sub>System</sub>* are the costs of purchased grid electricity without and with PV systems, respectively. *System Cost* is the initial investment cost of PV systems [6].

### 2.3.2. Payback period

Simple payback period (hereafter, payback period) is the time to recover the project cost of an investment, and can be expressed as the duration (e.g., years) from the initial investment to the time when the following condition is satisfied [38].

$$\sum_{n=1}^{t} \Delta I_n \leq \sum_{n=1}^{t} \Delta S_n$$

where $\Delta I$ is non discounted incremental investment costs ($) and $\Delta S$ is non discounted sum value of the annual cash flows net annual costs ($). $t$ represents the time when the condition is satisfied for the first time.

### 2.3.3. Levelized cost of electricity

Levelied cost of electricity (LCOE) is a measure of the average net present cost of PV electricity generation for its lifetime. It is useful to compare various sources of energy. LCOE ($/kWh) can be expressed as an following equation [30]:

$$LCOE = \frac{-C_0 - \sum_{n=1}^{N} \frac{C_n}{(1 + R_d)^n}}{\sum_{n=1}^{N} \frac{Q_n}{(1 + R_d)^n}}$$

where $C_0$ is the initial investment cost ($), $C_n$ is the annual project costs ($) in year $n$, and $Q_n$ is electricity (kWh) generated by the PV system in year $n$.

### 2.3.4. Environment and energy indicators

Analyzed results are also evaluated with following five environmental and energy indicators [6]. 1. Energy Sufficiency (ES) is how total PV generation can be compared to total annual demand. 2. Self-Sufficiency (SS) is how much PV generation can supply to local building demand considering hourly demand-supply balance. 3. Self-Consumption (SC) is how much PV generation can be consumed locally. 4. Cost Saving (CS) is how much energy costs can be saved by installing PV systems including capital and maintenance costs. 5. $CO_2$ emission reduction by PV systems is calculated compared with gird electricity consumption before the system installation. All the indicators are expressed in percentage. Equations of the relationships follow as:

1. ES = Total annual PV generation (kWh) / total annual demand (kWh) $\times 100$ (%)

2. SS = Total PV electricity amount locally consumed / total annual demand (kWh) $\times 100$ (%)

3. SC = Total PV electricity amount locally consumed / total annual PV generation (kWh) $\times 100$ (%)

4. Cost saving = (NPV/25) / (grid electricity cost)$_{base}$ $\times 100$ (%)

5. $CO_2$ emission reduction = {1 - ($CO_2$ emission from grid electricity consumption)$_{system}$ / ($CO_2$ emission from grid electricity consumption)$_{base}$} $\times 100$ (%) where $CO_2$ emission from grid electricity consumption = total imported grid electricity (kWh) $\times$ emission factor ($kg_{CO2}$ / kWh). Subscripts, "system" and "base" indicate building energy systems with "PV system" and "without PV system", respectively.

*2.4. Building scenarios*

To evaluate various building shapes in a block for energy demand and PV generation, we produced five scenarios (scenario 1-5) in comparison to existing buildings (scenario 0) (Fig. 6 and Table 2). Five scenarios are characterized by "Low-rise", "High-rise", "Center corridor", "Courtyard", and "Korean style". We set building widths in a range from 15 m to 50 m following general building shapes in the area. The "Low-rise" buildings (scenario 1) give pedestrian continuity along the front street. Therefore, they have advantages for small shops. "High-rise" buildings (scenario 2) are more independent to other buildings, which tends to foster unique identity to represent one company or residential buildings. Buildings with the "center corridor" (scenario 3) have a common open space between buildings. The open spaces around the buildings offer places for many community activities to workers, shops, and offices. "Courtyard" style (scenario 4) has similar shape as Scenario 3, but it is a typical courtyard type in Europe. Korean style (scenario 5) was adopted from building arrangements from Cheongye river, Seoul, South Korea, which is a well-known redeveloped waterfront area.

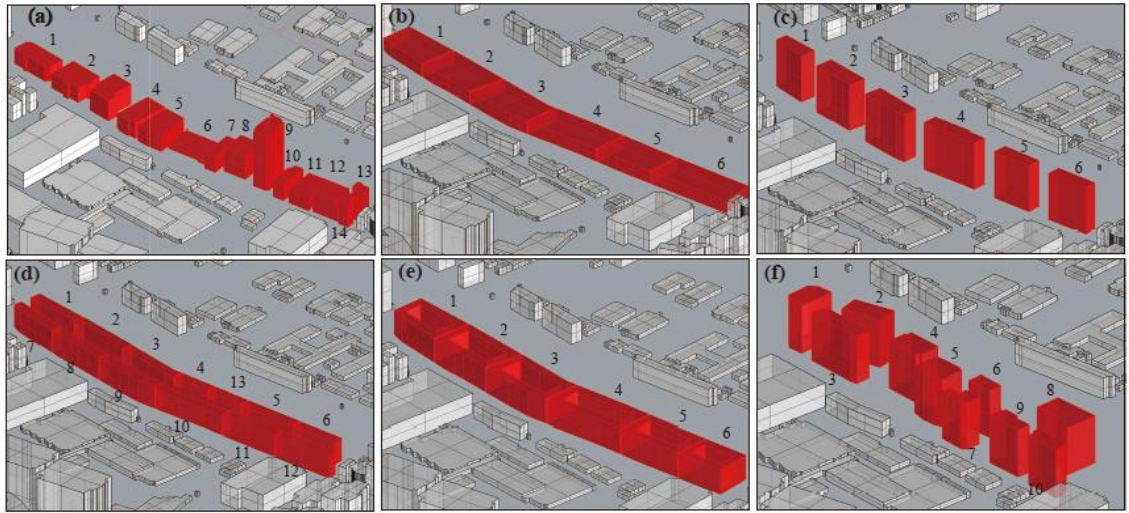

Figure 6. Existing buildings and five scenarios. (a) Scenario 0; existing buildings, (b) Scenario 1, (c) Scenario 2, (d) Scenario 3, (e) Scenario 4, and (f) Scenario 5. All the scenarios have the same FAR. Numbers next to buildings are building identification numbers.

In the following energy modeling, it is assumed that all the buildings in the scenarios are set to be used as "offices" for EnergyPlus. The analyses were conducted in an hourly resolution with weather information in 2018 for Shinagawa (Fig. 3). To compare various building morphology in comparison to existing buildings, floor area ratio (FAR) is set as a control variable. FAR is used to regulate building volumes and thus number of people in the districts or cities, which is inherently relate to necessary sizes of public services and goods such as water-sewer, road services, sun light availability, openness, and noises in the cities [39]. FAR is defined as:

$$FAR\ (\%) = \frac{Total\ floor\ area}{site\ area} \times 100 \quad\quad\quad (a)$$

As the site area (46,250 m$^2$) is common for all the scenarios, the total floor areas of buildings in the scenarios are also constant (Table 2). This results in total building volumes to be the same for all the scenarios (Table 2). Another important building indicator to regulate building forms, "building coverage ratio (BCR)" is defined as:

$$BCR\ (\%) = \frac{building\ area}{site\ area} \times 100 \quad\quad\quad (b)$$

Along with the FAR, BCR controls the shape and heights of buildings as well as occupied area by buildings in the site [40]. It is known that controlling BCR is an important policy measure to prevent spreading of fire. BCR varies between the scenarios from 18 % to 67 % (Table 2). Total surface areas above ground varies between the scenarios by 20 %. Surface area to volume ratio of buildings, which is an important indicator for energy balance of buildings, also varies by 20% among the scenarios (Table 2). Total rooftop areas vary by 46 % between the scenarios (Table 2). 70% of the total rooftop area is considered to be available for PV installation as a PV panel with efficiency of 20% need areas of about 5 m² plus an additional 2 m² for management or shaded areas, etc. Thus, 7 m²/kW is used as a coefficient to calculate maximum rooftop PV capacity for each scenario (Table 2).

Table 2. Characteristics of district scenarios. For the calculation of "surface area / volume", "above-ground surface area" was used. Numbers in parentheses were calculated by standard deviations divided by averages. See also figure 5 for 3D building representation of the scenarios. Bldg. is an abbreviation of building.

| Scenario # | 0 | 1 | 2 | 3 | 4 | 5 | Average | Standard deviation |
|---|---|---|---|---|---|---|---|---|
| Character | Existing | Low-rise | High-rise | Center corridor | Courtyard | Korean Style | | |
| Number of bldgs. | 14 | 6 | 6 | 13 | 6 | 10 | 9 | 4 (40%) |
| Average bldg. height | 27 | 18 | 66 | 33 | 33 | 54 | 39 | 18 (46%) |
| Average number of floors | 9 | 6 | 22 | 11 | 11 | 18 | 13 | 6 (46%) |
| Floor area (m²) | 185,000 | 185,000 | 185,000 | 185,000 | 185,000 | 185,000 | 185,000 | 0 (0%) |
| FAR (%) | 400 | 400 | 400 | 400 | 400 | 400 | 400 | 0 (0%) |
| BCR (%) | 44 | 67 | 18 | 36 | 36 | 22 | 37 | 17 (46%) |
| Total bldg. volume (m³) | 555,000 | 555,000 | 555,000 | 555,000 | 555,000 | 555,000 | 555,000 | 0 (0%) |
| Total surface area (m²) | 106,000 | 94,000 | 85,000 | 121,000 | 128,000 | 94,000 | 105,000 | 17,000 (16%) |
| Above-ground surface area (m$^2$) | 86,000 | 64,000 | 77,000 | 104,000 | 111,000 | 84,000 | 87,000 | 17,000 (20%) |
| Surface area (m$^2$) / volume (m$^3$) | 0.15 | 0.11 | 0.14 | 0.19 | 0.20 | 0.15 | 0.16 | 0.03 (20%) |
| Total rooftop area (m$^2$) | 20,200 | 30,800 | 8,400 | 16,800 | 16,800 | 10,300 | 17,200 | 8,000 (46%) |
| Total rooftop PV capacity (MW) | 2.89 | 4.40 | 1.20 | 2.40 | 2.40 | 1.47 | 2.46 | 1.14 (46%) |

## 3. Results

### 3.1. Building energy demands

Estimated energy demands for buildings include interior lighting, interior electric equipment, space heating, and space cooling with typical office use activity in an hourly resolution. Interior equipment consumes the largest amount of electricity by 68% of the total, and lighting is 12% (Table 3). In addition, space heating and cooling are 2 % and 18 %, respectively. Demands for lighting and interior equipment are constant among the scenarios (Table 3) as they are generally functions of floor area. Space heating, which shows the largest variability (13 % of average) between the scenarios, has significant correlation with surface area/volume ratio, explaining 97% of variance (Fig. 7). Space cooling, although much smaller variability (0.8 % of average), has significant correction with building height or number of floors (Table 3, Fig. 7). Little variability of space cooling among the scenarios indicate that floor space or volume of building (set constant in this study) is the most important factor, and surface area-volume ratio has little impacts on space cooling. Unit floor electricity consumptions (kWh/m$^2$) are generally consistent with available data of 290 kWh/m$^2$ for office building electricity consumption for 2019-2020 [41]. Slightly older data of 2014 shows a larger average value of 389 kWh/m$^2$ for office buildings in Kanto area [42].

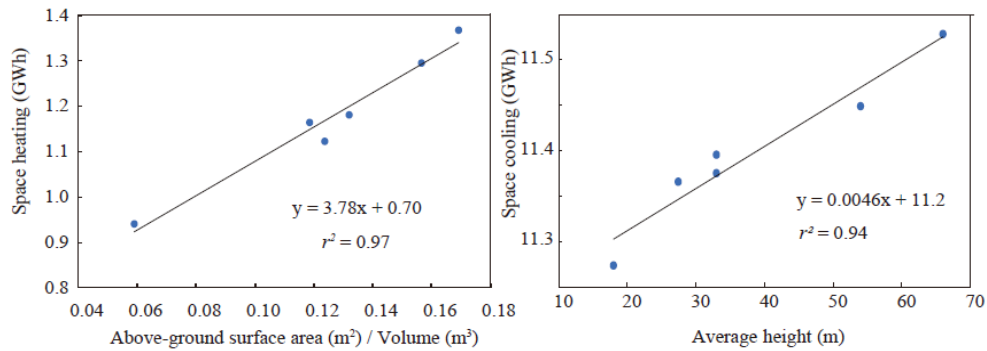

Figures 7. Influences on space heating and cooling.

Space cooling is the second largest demand and shows clear relationship with outside temperatures from April to November (Fig. 3). On the other hand, demand for space heating is so small that its correlation with out-side temperature in winter is not clear (Fig. 3). Total energy demands show clear weekly cycles (national holidays are not considered) (Fig. 3). Lighting and interior equipment do not have seasonal changes. Estimated hourly total demands for all the scenarios are highly correlated ($r > 0.99$) with that of the existing buildings (scenarios 0) as high demand components (i.e., interior equipment, lighting, and space cooling) do not show differences between the scenarios.

Table 3. Calculated building energy demands for the scenarios. Parentheses at average section represent percentages of demand components to the total demand. Parentheses at standard deviation section represent percentages of standard deviation divided by averages.

| Scenario # | 0 | 1 | 2 | 3 | 4 | 5 | Average | Standard deviation |
|---|---|---|---|---|---|---|---|---|
| Name | Existing | Low-rise | High-rise | Center-corridor | Courtyard | Korean Style | | |
| Lighting (GWh) | 7.7 | 7.7 | 7.7 | 7.7 | 7.7 | 7.7 | 7.7 (12%) | 0 (0%) |
| Interior Equipment (GWh) | 42.2 | 42.1 | 42.2 | 42.2 | 42.2 | 42.2 | 42.2 (68%) | 0 (0%) |
| Space heating (GWh) | 1.2 | 0.9 | 1.1 | 1.3 | 1.4 | 1.2 | 1.2 (2%) | 0.2 (13%) |
| Space cooling (GWh) | 11.3 | 11.2 | 11.5 | 11.3 | 11.3 | 11.4 | 11.3 (18%) | 0.1 (0.8%) |
| Total consumption electricity (GWh) | 62.4 | 62.0 | 62.5 | 62.6 | 62.6 | 62.5 | 62.4 | 0.2 (0.3%) |
| Unit floor electricity consumption (kWh/m$^2$) | 337 | 336 | 338 | 338 | 338 | 338 | 337 | 1.0 (0.3%) |

To investigate the impacts of shading, we conducted a test "with shades" and "without shades" for the building # 8 in the existing buildings (scenario 0; Fig. 6). The shades were produced by the building # 9, which is the tallest building in all the buildings considered and located south of the building # 8 (Fig. 6). The analysis was conducted in the same way for the scenario analyses for one year in an hourly resolution. Results show that the influence of shades are negligible in terms of the total annual energy demand (Table 4). However, the demand for space heating increases by shades by 5.7%, but the demand for space cooling reduces by 0.5% (Table 4). As absolute numbers are similar between demands for space heating and cooling but with opposite signs, they cancel each other with little change in the total (Table 4). Therefore, we conclude that shading has little influence on the total building energy demands in the settings we considered.

Table 4. Influences of shades on building energy demand of building #8 of scenario 0. Parentheses for "shaded" and "no shade" are ratios to the total. Parentheses for "difference" are ratios with "no shade". "Difference" is calculated as "Shaded" – "No shade".

| | Shaded (%) | No shade (%) | Difference (%) |
|---|---|---|---|
| Space heating (MWh) | 104.3 (1.8) | 98.7 (1.7) | 5.6 (5.7%) |
| Space cooling (MWh) | 1,065.4 (18.2) | 1,071.2 (18.3) | -5.8 (-0.5%) |
| Interior lighting (MWh) | 725.7 (12.4) | 725.7 (12.4) | 0 (0.0%) |
| Interior equipment (MWh) | 3,955.3 (67.6) | 3,955.3 (67.6) | 0 (0.0%) |
| Total (MWh) | 5,850.7 (100.0) | 5,850.9 (100.0) | -0.2 (0.0%) |

*3.2. Technoeconomic analysis with rooftop PV systems*

Building integrated PV (BIPV) are increasingly important for urban decarbonization when costs of PV systems are declining and land areas for PV are limited such as for Japan. However, BIPV potential for buildings in urban environment are affected by building forms and relationships between neighboring buildings for available sunlight (Fig. 4). Analyses on existing buildings show that 81% of the rooftop area can receive more than 90% of the maximum solar radiation (Table 5). On the other hand, southern faced façade with no shades receives only 60-70 % of the maximum solar radiation. As the cost of installation of rooftop PV systems are lower than that for facades [43], the first priority is given to the rooftop, but newly developing PV materials with light weight such as perovskite PV could allow future application on façade more economic and easier than at present. All other scenarios have constant heights for all the buildings such that the rooftops of these buildings receive the maximum solar radiation.

Table 5. Radiation analysis on the existing buildings (scenario 0; Fig. 4). Top and bottom represent a range of annual radiation received for mesh. "Above max radiation" indicates percentage of the maximum annual radiation (1,383 kWh/m$^2$). Rooftop and façade areas are percentages of the total areas. For example, 81% of the rooftop area received more than 90 % of the maximum annual radiation.

| Top (kWh/m$^2$) | Bottom (kWh/m$^2$) | Above max radiation (%) | Rooftop area (%) | Façade area (%) |
|---|---|---|---|---|
| 1,383 | 1,244 | 90 | 81 | 0 |
| 1,244 | 1,106 | 80 | 91 | 0 |
| 1,106 | 968 | 70 | 96 | 0 |
| 968 | 830 | 60 | 98 | 10 |
| 830 | 691 | 50 | 100 | 19 |
| 691 | 553 | 40 | 100 | 51 |
| 553 | 415 | 30 | 100 | 56 |
| 415 | 277 | 20 | 100 | 60 |
| 277 | 138 | 10 | 100 | 78 |
| 138 | 0 | 0 | 100 | 100 |

As total rooftop areas are variable between the scenarios, amounts of PV generation is highly variable (Table 2). "Low-rise" buildings (scenario 1) have the largest rooftop area and PV capacity (4.4 MW) installed. Annual PV generation is 5.53 GWh, which supplies 8.9% of demand (Table 5). As the PV generation is small compared with the building demands, all PV generated electricity is consumed on site (100 % self-consumption) for all the scenarios (Table 6). Thus, self-sufficiency and energy sufficiency are equal. In addition, as $CO_2$ emission reduction is equal to the amount of grid electricity replaced by PV electricity, the values for $CO_2$ emission reduction are the same with self- and energy-sufficiency.

If PV are installed on the façade receiving >50 % (691 kWh/m$^2$) of the maximum solar insolation for the existing buildings, the corresponding façade area is 12,425 m$^2$ (19% of the total façade area) and the façade PV is expected to generate annually 2.1 GWh of electricity, which is about the same amount with the annual rooftop PV generation, emphasizing the need to utilize the façade to further increase on-site PV generation.

Table 6. Energy indicators for 2018 and 2030. Results are the same for both years as optimal PV capacities are the same for both years. 70% of the rooftop area is used for the maximum PV installation.

| 2018 and 2030 | S0 | S1 | S2 | S3 | S4 | S5 |
|---|---|---|---|---|---|---|
| Rooftop PV generation (GWh/yr) | 3.63 | 5.53 | 1.51 | 3.02 | 3.02 | 1.85 |
| Self-sufficiency | 5.8% | 8.9% | 2.4% | 4.8% | 4.8% | 3.0% |
| Self-consumption | 100% | 100% | 100% | 100% | 100% | 100% |
| Energy sufficiency | 5.8% | 8.9% | 2.4% | 4.8% | 4.8% | 3.0% |
| $CO_2$ emission reduction | 5.8% | 8.9% | 2.4% | 4.8% | 4.8% | 3.0% |

Rapidly declining cost of PV system creates increasing opportunities for affordable decarbonization of buildings with BIPV. For 2018, LCOE of 0.13 $/kWh is slightly lower than high-voltage electricity price of 0.15 $/kWh, indicating that rooftop PV is already economic in 2018. As self-consumption is 100% for all the scenarios, simple payback period are the same values for all the scenarios. Payback period is 14 years in 2018, which is still longer than generally considered promising investment opportunities [44]. Cost saving is less than 1 %, which indicates difficulty to motivate owners to invest on on-site PV systems in 2018. By 2030, economic situations of rooftop PV for these buildings improve significantly. LCOE becomes 0.07$/kWh, which is half of the tariff price. Payback period of 6 years is less than half of that of 2018. However, cost saving for energy expense is only 1-3 % owing smaller PV generation in comparison to the demand. NPVs in 2030 become nearly four times larger than that in 2018. The maximum NPV reaches $7.0 million for the low-rise buildings (scenario 1), which is 3.7 times larger than that for the high-rise building (Table 7).

Table 7. Financial indicators of rooftop PV systems for 2018 and 2030.

| 2018 | S0 | S1 | S2 | S3 | S4 | S5 |
|---|---|---|---|---|---|---|
| LCOE ($/kWh) | 0.13 | 0.13 | 0.13 | 0.13 | 0.13 | 0.13 |
| NPV (million $) | 1.21 | 1.85 | 0.50 | 1.00 | 1.00 | 0.62 |
| Payback period (yrs.) | 14 | 14 | 14 | 14 | 14 | 14 |
| Cost saving | 0.5% | 0.8% | 0.2% | 0.4% | 0.4% | 0.3% |
| 2030 | S0 | S1 | S2 | S3 | S4 | S5 |
| LCOE ($/kWh) | 0.07 | 0.07 | 0.07 | 0.07 | 0.07 | 0.07 |
| NPV (million $) | 4.60 | 7.00 | 1.91 | 3.82 | 3.82 | 2.33 |
| Payback period (yrs.) | 6 | 6 | 6 | 6 | 6 | 6 |
| Cost saving | 2.0% | 3.0% | 0.8% | 1.6% | 1.6% | 1.0% |

## 4. Discussions

In fall 2020, Japanese government declared a goal to reach carbon neutral by 2050, which substantially changed social atmosphere toward carbon neutrality. Primary sources of carbon-free energy will be renewable energies, in particular, "solar power" [45]. As available lands for PV installation are limited in Japan and for the sake of saving natural lands [46], it is critical to maximize the rooftop uses for PV generation in a physically maximum extent. Although feed-in-tariffs (FITs) helped successfully expand PV in Japan [47], rooftop PVs of large buildings have received little attention as rooftop PV generation is rather small compared to large building demands. As our analyses indicated, declining cost of PV systems for the next decades will create large economic benefits to install PV on the rooftops of large buildings. In addition, total rooftop areas of large buildings in cities are not negligible and difficult to be replaced on the ground. Therefore, it is important to place adequate policy measures to facilitate expansion of rooftop PV systems on large buildings.

A few relevant policies can be recommended in this regard. First, current building form regulations mainly through FAR and BCR could be upgraded to promote the installation of rooftop PV on new buildings. One measure is to introduce a performance-based regulation under the designated FAR where energy efficiency is set as one of the core performances mandated to buildings. Another measure is to craft a form-based regulation designating wall setbacks, maximum height, minimum "Roof Area Ratio", etc. to facilitate building forms that allow effective rooftop PV installation. Second, in Japan where population and economy are shrinking in many cities, it is possible to reduce the designated FAR in existing urban area, which will prevent the construction of high-rise buildings that have less opportunities to install rooftop PV. But this measure should be carefully discussed in relation to the compact city-plus-network concept that promotes high density urban area to reduce energy consumption in transport sector.

Toward carbon neutrality, critical measure for buildings are energy efficiency improvements including electrification [48]. Lighting, electric equipment, space cooling and heating have large potentials to reduce energy consumption but providing the same or better services by switching to more efficient and smart apparatus [49]. Passive solutions also play important roles particularly during building design phases, which are, for example, building materials for thermal insulation, building forms, window-to-wall ratio leading to reduced energy consumption. However, building design should be considered not solely from the energy efficiency requirement but also from other perspectives including the livability inside the building and the quality of environment outside building [12]. Buildings also need to prepare for adaptation to increasing climate change by enhancing resilience, etc., which often overlaps with mitigation measures such as developing decentralized energy systems [50].

In a city scale, rooftop PVs coupled with electric vehicles (EV) are possibly highly effective tools to decarbonize urban energy systems [5,6]. EVs are important means to decarbonize transport sector by replacing internal combustion of fossil fuels with electricity, but also play roles as energy storage for VRE such as rooftop PVs [25] or wind power [51]. It has been shown that rooftop PVs combined with EVs in a city scale can supply up to 95 % of electricity to cities in Japan [6]. In the case of the special districts of Tokyo including Shinagawa, the PV plus EV systems can supply 53% of the annual demands. As large buildings of the central urban area such as this study consumes all the on-site PV generation, EVs or batteries have no roles to play as energy storage. However, Tokyo has various types of districts with independent houses and small buildings with smaller demands. Rooftop PVs of these buildings produce excess electricity, which supplies part of the demands of the large buildings by coupling with EVs. To fully understand the supply and demand balance of urban power systems, it is necessary to analyze disaggregated rooftop PV generation coupled with EVs within cities considering grid constrains.

## 5. Conclusions

This study established a workflow of assessing waterfront office building redevelopment plans with rooftop photovoltaics (PV), different building shapes and arrangements. We produced five scenarios in comparison to existing buildings with the same floor area ratio (FAR) and total floor areas. Demands for space heating are found to strongly correlate with surface area to volume ratio, although space heating demand for buildings are small in comparison to the total demand of buildings for the cases we considered. Shades by neighboring buildings affect space heating and cooling demands in opposite signs and cancel each other. Therefore, shades have little influence on the total energy demand for the buildings. Rooftop PV is already economic in 2018, and by 2030 it improves substantially with payback periods reaching 6 years. However, the rooftop PV contribution on the total demands of buildings or $CO_2$ emission reduction is small (2-9%), as rooftop areas are limited on the large buildings. However, places like Japan where lands are limited for PV installation should implement mandatory regulation of installing rooftop PVs on buildings as such investments should not be issues for building owners. In addition, it is recommended to upgrade building form regulation to promote the installation of rooftop PV on new buildings.

**CRediT authorship contribution statement**

**Choi Younghun:** Conceptualization, methodology, Investigation, Writing – Original draft. **Takuro Kobashi:** Conceptualization, Methodology, Validation, Formal analysis, Investigation, Software, Writing - original draft, Writing - review & editing, Visualization, Supervision. **Yoshiki Yamagata:** Funding acquisition. **Akito Murayama**: Writing – Original Draft.

**Declaration of competing interest**

The authors declare that they have no known competing financial interests or personal relationships that could have appeared to influence the work reported in this paper.

**Acknowledgements**

We appreciate Dr. Soowon Chang at Purdue University for sharing Grasshopper files for the building energy analyses.

**Supplementary data**

Supplementary data to this article can be found online at

https://doi.org/10.17632/wfpkdc6rd7.1